%% file: main_arx.tex
\documentclass[pra,superscriptaddress,noeprint,twocolumn]{revtex4-2}
\usepackage{amsmath}
\usepackage{amsfonts}       
\usepackage{graphicx}
\usepackage{dcolumn}
\usepackage{epsfig}
\usepackage[mathscr]{eucal}
\usepackage{bm}
\usepackage{array}
\usepackage{hyperref}
\hypersetup{
colorlinks=true,
citecolor=blue,
linkcolor=blue,
urlcolor=blue,
}
\usepackage[dvipsnames]{xcolor}			
\usepackage{tikz}

\usepackage{amssymb}
\usepackage{bm}           
\usepackage{bbm}
\usepackage[english]{babel}
\usepackage{calrsfs}	
\usepackage{dutchcal}
\DeclareMathAlphabet{\pazocal}{OMS}{zplm}{m}{n}	

\definecolor{DeepSkyBlue}{RGB}{0,104,139}
\colorlet{MySky}{white!40!blue}
\colorlet{MyViolet}{red!45!blue}
\colorlet{MyBlue}{black!40!blue}
\colorlet{MyRed}{black!40!red}
\colorlet{MyOrange}{red!70!yellow}
\colorlet{MyGreen}{black!60!green}
\colorlet{MyBrown}{black!70!brown}
\colorlet{MyGray}{black!60!white}

\newcommand{\be}{\begin{equation}}
\newcommand{\ee}{\end{equation}}

\newcommand{\beq}{\begin{eqnarray}}
\newcommand{\eeq}{\end{eqnarray}}

\newcommand{\bra}[1]{\ensuremath{\langle#1|}}
\newcommand{\ket}[1]{\ensuremath{|#1\rangle}}

\begin{document}

\title{Beyond Pure Dephasing: Quantum Error Correction in \\
Single Molecules Requires Multiple Spins}

\author{Matteo Mezzadri}
\thanks{These authors contributed equally to this work.}
\affiliation{Dipartimento di Scienze Matematiche, Fisiche e Informatiche, Universit\`a  di Parma, Parco Area delle Scienze, 53/A, I-43124 Parma, Italy.}
\affiliation{Gruppo Collegato di Parma, INFN-Sezione Milano-Bicocca, I-43124 Parma, Italy.}

\author{Silvia Macedonio}
\thanks{These authors contributed equally to this work.}
\affiliation{Dipartimento di Scienze Matematiche, Fisiche e Informatiche, Universit\`a  di Parma, Parco Area delle Scienze, 53/A, I-43124 Parma, Italy.}
\affiliation{Gruppo Collegato di Parma, INFN-Sezione Milano-Bicocca, I-43124 Parma, Italy.}

\author{Luca Lepori}
\thanks{These authors contributed equally to this work.}
\affiliation{Dipartimento di Scienze Matematiche, Fisiche e Informatiche, Universit\`a  di Parma, Parco Area delle Scienze, 53/A, I-43124 Parma, Italy.}
\affiliation{UdR Parma, INSTM, I-43124 Parma, Italy.}
\affiliation{Gruppo Collegato di Parma, INFN-Sezione Milano-Bicocca, I-43124 Parma, Italy.}
\affiliation{Dipartimento di Fisica, Universit\`a di Pisa, Largo Pontecorvo 3, I-56127 Pisa, Italy.}

\author{Emilio Macaluso}
\affiliation{Dipartimento di Scienze Matematiche, Fisiche e Informatiche, Universit\`a  di Parma, Parco Area delle Scienze, 53/A, I-43124 Parma, Italy.}
\affiliation{Gruppo Collegato di Parma, INFN-Sezione Milano-Bicocca, I-43124 Parma, Italy.}
\affiliation{UdR Parma, INSTM, I-43124 Parma, Italy.}

\author{Francesco Albarelli}
\affiliation{Dipartimento di Scienze Matematiche, Fisiche e Informatiche, Universit\`a  di Parma, Parco Area delle Scienze, 53/A, I-43124 Parma, Italy.}
\affiliation{Gruppo Collegato di Parma, INFN-Sezione Milano-Bicocca, I-43124 Parma, Italy.}

\author{Richard E.P. Winpenny}
\affiliation{Department of Chemistry and Photon Science Institute, The University of Manchester, Oxford Road, Manchester M13 9PL, United Kingdom.}

\author{Alessandro Chiesa}
\email{alessandro.chiesa@unipr.it}
\affiliation{Dipartimento di Scienze Matematiche, Fisiche e Informatiche, Universit\`a  di Parma, Parco Area delle Scienze, 53/A, I-43124 Parma, Italy.}
\affiliation{Gruppo Collegato di Parma, INFN-Sezione Milano-Bicocca, I-43124 Parma, Italy.}
\affiliation{UdR Parma, INSTM, I-43124 Parma, Italy.}

\author{Stefano Carretta}
\email{stefano.carretta@unipr.it}
\affiliation{Dipartimento di Scienze Matematiche, Fisiche e Informatiche, Universit\`a  di Parma, Parco Area delle Scienze, 53/A, I-43124 Parma, Italy.}
\affiliation{Gruppo Collegato di Parma, INFN-Sezione Milano-Bicocca, I-43124 Parma, Italy.}
\affiliation{UdR Parma, INSTM, I-43124 Parma, Italy.}

\begin{abstract}
We propose multi-spin molecules as a viable architecture for fault-tolerant quantum computing.
To this aim, we introduce a correction protocol handling both diagonal and off-diagonal errors, typically associated with dephasing and relaxation. The scheme is based on a hybrid encoding which combines dephasing-tolerant units  suppressing the leading pure dephasing error into a multi-spin molecule implementing a multi-qubit code for residual off-diagonal errors. 
Our proposal leverages peculiar properties of molecular spins, i.e. the strong hierarchy between different errors and the possibility to engineer multi-spin molecules at the synthetic level. Moreover, it addresses
the important issue of the loss of coherences in anharmonic systems subject to off-diagonal errors, which hampers their correction at the single spin level. 
Thanks to the huge suppression of dephasing by the first-level code, we numerically demonstrate the potential performance of this strategy even with a limited number of spins per logical unit.
\end{abstract}

\maketitle
\section{Introduction}

The practical implementation of quantum error correction (QEC) is essential for realizing the full potential of quantum computing.
Furthermore, QEC must be implemented in a fault-tolerant (FT) manner, keeping the errors under control at all stages, to realise universal quantum computation in the real world~\cite{nielsen, KLC, QEC_for_general_noise, threshold_theorem, review_surface_code_2012}.
Qudits are particularly promising for embedding QEC against specific errors with low overhead~\cite{GKP, Qudit_Camerino, Binomial_Codes_Girvin,roffe2019,mezzadri2024}.

Embedding QEC codes in single qudits has been investigated in bosonic platforms~\cite{First_Bosonic_Code, GKP, Qudit_Camerino, Binomial_Codes_Girvin},  molecular spins~\cite{Chiesa2020_JPCL_prima_proposta_EQEC_MNM, Petiziol_optimized_encoding, Chiesa_QEC_2022,mezzadri2024,mezzadri2025}, and cold atoms \cite{deutsch2024}. 
Considerable effort has been devoted to QEC codes targeting dephasing errors~\cite{Chiesa2020_JPCL_prima_proposta_EQEC_MNM, Petiziol_optimized_encoding, Chiesa_QEC_2022,mezzadri2024}, i.e. noise processes described by operators diagonal in the energy eigenbasis.
The choice to correct dephasing first is motivated by a strong hierarchy of decoherence channels in systems such as molecular nanomagnets (MNMs)~\cite{chiesa2024} and other spin systems~\cite{Fuentes2024,Yu2025}, in which off-diagonal errors occur on much longer timescales than pure dephasing.
Molecular spins such as MNMs are particularly appealing because they typically have the eigenstate connectivity needed for FT implementations of embedded codes against diagonal errors~\cite{mezzadri2024}.
Recently, experimental steps toward the realization of embedded QEC in MNMs have been demonstrated~\cite{ardavan2024}, and an application to quantum sensing with sensitivity enhanced by QEC has been proposed~\cite{mezzadri2025}.

Once the leading dephasing error is corrected, however, slower non-diagonal errors become the bottleneck for FT computation.
Here we analyze the requirements for FT correction of such errors in qudit architectures characterized by sizable differences between energy gaps (anharmonicity), such as MNMs~\cite{chiesa2024}.
In this regime, the \emph{secular approximation} resolves relaxation into independent channels associated with distinct transitions, often of comparable strength, even if the spin–bath coupling is modeled by \emph{global} spin operators acting on the whole qudit space.
This goes beyond spin-qudit encodings that take only a small set of global spin operators as the error model to be corrected~\cite{Gross2021a,Lim2023,Lim2025,Omanakuttan2023a,deutsch2024,Lim2026}.
Crucially, the resulting loss of transferred coherence challenges the feasibility of resource-efficient QEC within a single spin~\cite{albert2018,albert2024,furey2024}.

We address this issue by proposing multi-spin molecules as the route to embed QEC against generic dephasing and relaxation (diagonal and non-diagonal) errors into single physical units. 
We exploit the possibility of tailoring chemical structures to realize multi-spin clusters supporting a two-level encoding: pure dephasing errors are corrected within spin $S-\mathcal{s}$ dimers~\cite{mezzadri2024} (with $S\geq3/2$) and the resulting dephasing-tolerant units are concatenated in a multi-spin molecule  to handle residual errors by a second-level multi-qubit code.
The advantage gained at this first level is then amplified by concatenation, since the overall logical error scales polynomially with the residual first-level error.
Thus, an improvement at the first level can translate into orders-of-magnitude stronger logical-error suppression, or a smaller outer code for a fixed target accuracy.
Our estimates indicate substantial gains over standard multi-qubit codes even with a limited number of spins, suggesting molecular spins as a hardware-efficient route toward fault-tolerant quantum computing.
Application to other spin systems such as neutral atoms is finally discussed. 

The manuscript is organised as follows: we first review our approach to develop dephasing-tolerant units consisting of $S-\mathcal{s}$ spin pairs. We then address the correction of off-diagonal (relaxation) errors, showing the impossibility of correcting them with a single-spin encoding. We finally propose a solution based on multi-spin clusters and show its effectiveness in suppressing both dephasing and relaxation.

\section{Results}

\subsection{Dephasing-tolerant Molecular Spin Qudits}
We first review embedded correction of pure dephasing in molecular spin qudits~\cite{AIPadv, Chiesa2020_JPCL_prima_proposta_EQEC_MNM, Chiesa_QEC_2022,ChizziniPCCP,mezzadri2024}, which provides the first-level dephasing-tolerant  units for our two-level encoding.

We consider the spin Hamiltonian
\begin{equation}
    \label{eq:Ham}
    H_S = g_z \mu_B B_z S_z + D\,S_z^2 + E \, (S_x^2 - S_y^2) \, ,
\end{equation}
which describes a single-ion or multi-spin single-molecule magnet with ground total spin $S\geq3/2$.
The first term is the Zeeman interaction with a magnetic field along $z$, while $D$ and $E$ parameterize the axial and rhombic zero-field splitting.
This Hamiltonian is typical of many molecular spins~\cite{Freedman_Cr,Jenkins2017,SIMqubit}.
Our simulations use $B_z=0.35$~T, with $D=-0.81$~cm$^{-1}$ and $E=-0.24$~cm$^{-1}$ for $S=3/2$~\cite{Freedman_Cr}, and $D=0.0427$~cm$^{-1}$ and $E=0.0098$~cm$^{-1}$ for $S=7/2$~\cite{Jenkins2017}.
Nuclear spin qudits in MNMs, coupled to electronic-spin ancillae for error detection, are also possible candidates for embedded QEC~\cite{Chiesa2020_JPCL_prima_proposta_EQEC_MNM,ChizziniPCCP,ardavan2024}.

The diagonal spin--bath coupling yields the master equation~\cite{Chiesa_QEC_2022,PhysRevResearch.4.043135}
\begin{equation}
\dot{\rho}(t)  = - \sum_{\mu \mu'} \gamma_{\mu \mu'} \bra{\mu} \rho \ket{\mu'} \, \ket{\mu}\bra{\mu'}  \, ,
\label{eq:MEq_diag}
\end{equation}
where $\rho$ is the density operator, $\ket{\mu}$ are eigenstates of $H_S$, and
\begin{align} \nonumber
    \gamma_{\mu\mu'} &= \frac{1}{T_2} \bigg( \bra{\mu} S_z \ket{\mu} \bra{\mu} S_z \ket{\mu} + \bra{\mu'} S_z \ket{\mu'} \bra{\mu'} S_z \ket{\mu'} \\
    &-2 \bra{\mu} S_z \ket{\mu} \bra{\mu'} S_z \ket{\mu'} \bigg) \, ,
\end{align}
with $T_2$ the characteristic dephasing time.
This independent decay of the off-diagonal elements of $\rho$ was also confirmed by simulations beyond the Markov approximation using the cluster-correlation expansion~\cite{Petiziol_optimized_encoding,Ratini2025}.

At a fixed short time $t$, the evolution can be expressed through the operator-sum representation
\begin{equation}
    \rho(t)=\sum_k E_k\,\rho(0)\,E_k^\dagger \, ,
    \label{rhoev}
\end{equation}
with at most $d^2$ Kraus operators for a $d$-level qudit.
In practice, this can be obtained via process tomography~\cite{nielsen}. 
For pure dephasing, these operators are diagonal and typically exhibit a hierarchy of relevance~\cite{Chiesa2020_JPCL_prima_proposta_EQEC_MNM,mezzadri2024}.
For the embedded codes considered here, we select the $d/2$ leading operators and seek codewords $\{\ket{0_L},\ket{1_L}\}$ satisfying the Knill--Laflamme conditions~\cite{KLC}:
\begin{eqnarray}
    \label{eq:KL}
    \bra{0_L}E_k^\dagger E_j\ket{0_L}
    &=& \bra{1_L}E_k^\dagger E_j\ket{1_L} \, ,\\
    \bra{0_L}E_k^\dagger E_j\ket{1_L}
    &=& 0 \, . \nonumber
\end{eqnarray}
Gram--Schmidt orthogonalization of
$\{E_k\ket{0_L},E_k\ket{1_L}\}_{k=0}^{d/2-1}$
then gives an orthonormal basis $\{\ket{\ell,k}\}$, where $\ell=0,1$ labels the logical states and $k=0,\ldots,d/2-1$ labels the correctable errors.
The Knill--Laflamme conditions ensure that these errors preserve the encoded logical information.

To achieve effective QEC, codewords must be complemented by a FT implementation of logical gates and correction.
Several spin-qudit proposals address aspects of this requirement through logical operations, pulse sequences, or gate-fidelity estimates~\cite{Gross2021a,Omanakuttan2023a,Lim2023,Lim2026}, while a complete FT analysis remains a separate step.
Here, we consider an error-transparent implementation, such that logical gates act identically in every error subspace $k$, without mixing different subspaces~\cite{Ma2020,mezzadri2024,mezzadri2025}.
This construction exploits  suitable \emph{connectivity} between eigenstates, i.e. the ability to drive direct resonant transitions between them~\cite{mezzadri2024}.
Moreover, FT error detection and recovery also exploit this connectivity~\cite{mezzadri2024}.
A $d/2$-level ancilla, initialized in its ground state and coupled to the $d$-level qudit, is acted upon by an operation controlled by the error label $k$.
Reading out the ancilla in its eigenbasis projects the qudit onto the corresponding error subspace, which is then mapped back to the code space.
We refer to the resulting $S-\mathcal{s}$ dimer, with $S=(d-1)/2$ hosting the logical qubit and $\mathcal{s}=(d/2-1)/2$ the ancilla, as a \emph{dephasing-tolerant qubit}.

\begin{figure}[ht!]
\centering
\includegraphics[width=0.9\linewidth]{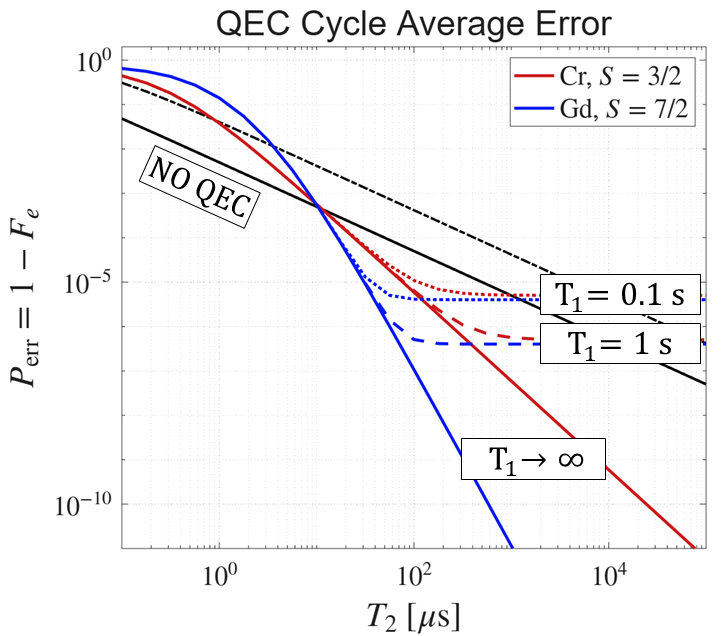}
 \caption{Average error of a logical rotation followed by syndrome extraction and recovery, as a function of the dephasing time $T_2$.
The error is quantified by the entanglement infidelity
$P_{\rm err}=1-F_e$ of the logical channel with respect to the target unitary and is averaged over the universal set of planar logical rotations
$R_L^p(\theta,\phi)=\exp[-i(\cos\phi\,Y_L-\sin\phi\,X_L)\theta/2]$,
with $(\theta,\phi)=(\pi/4,\pi)$, $(\pi/2,\pi)$,
$(\pi/2,-\pi/2)$, $(\pi/2,-\pi/4)$, and
$(\pi/2,-\pi/8)$.
Red and blue curves correspond to the $S=3/2$ Cr complex \cite{Freedman_Cr}
and to the $S=7/2$ Gd complex \cite{Jenkins2017}, respectively.
Solid, dashed, and dotted curves correspond to
$T_1\rightarrow\infty$, $T_1=1$~s, and $T_1=100$~ms, respectively.
For each qudit, the duration of each operation is set by the smallest dipole matrix element among the transitions involved in the logical gate, syndrome extraction, and recovery, taking \(10\) ns as the reference duration for a $\pi$ pulse on a spin 1/2 \cite{hiper}.
This yields 
total gate--syndrome--recovery sequence durations of $250$~ns and $551$~ns, for $S=3/2$ and $S=7/2$, respectively.
The black curves show an
uncorrected $S=1/2$ subject to the same dephasing (no relaxation) for the time required by a logical gate on the spin 3/2 ($\sim 83$ ns, dashed) or by the manipulation of a bare spin 1/2 (10 ns, solid).
In the absence of relaxation, increasing the spin size leads to a progressively stronger suppression of dephasing errors, whereas for finite $T_1$ the curves eventually saturate to a $T_2$-independent plateau set by residual relaxation.}
    \label{fig:confronto_T1}
\end{figure}

Figure~\ref{fig:confronto_T1} shows the logical error (quantified by the entanglement infidelity, see SI for a precise definition and operational
interpretation) after a rotation followed by a QEC cycle, obtained by simulating the pulse sequence under continuous dephasing.
The simulation details are given in the caption.
Compared with an uncorrected spin-$1/2$ system undergoing the same dephasing, the embedded codes exhibit an impressive decrease of the error with $T_2$, with stronger suppression for larger spin.
This behavior witnesses the FT implementation described above, possible thanks to the all-to-all connectivity of $H_S$ in Eq.~\eqref{eq:Ham}.

Even though $T_1\gg T_2$ in the materials considered here, including relaxation in the simulations produces an error floor, as shown by the dashed and dotted curves in Fig.~\ref{fig:confronto_T1}.
Without targeted correction of relaxation, this floor increases as $T_1$ decreases and limits the benefit of suppressing dephasing.
We thus examine how to correct relaxation errors within an embedded encoding.
To this aim, we start by reviewing their open-system dynamics.

\subsection{The issue of non-diagonal errors}
We assume the spin system is coupled to a bosonic bath through an  operator of the form
\begin{equation}
    V(t) = \sum_i e^{i(H_S+H_B)t/\hbar} Q_i F_i e^{-i(H_S+H_B)t/\hbar} ,
    \label{eq:hamSB}
\end{equation}
where $F_i$ are bath operators, $Q_i$ are operators acting only on the system and we work in the interaction picture with respect to the uncoupled system and bath Hamiltonians $H_S$ and $H_B$.
With general assumptions \cite{Blum,BPO,Vacchini2024} we can derive the Redfield master equation for the time evolution of the system density matrix $\rho$ in the basis of the eigenstates $\ket{m}$ of $H_S$:
\begin{equation}
\label{eq:masteq}
    \bra{m'}\dot{\rho}  \ket{m} = 
     \sum_{n',n}\bra{n'}\rho  \ket{n} R_{m'mn'n} \, e^{ i \, (\Delta_{m' n'}-\Delta_{m n}) \, t/\hbar} .
\end{equation}
Here $\Delta_{mn}=(\varepsilon_{m}-\varepsilon_n)$  are the energy gaps between the eigenstates $\ket{m}$ and  $\ket{n}$, and $R_{m'mn'n}$ is the Redfield tensor. To draw general conclusions independently of the details of the bath, we assume a wide-band limit for the bath spectral function and get
\begin{align}\nonumber
   & R_{m'mn'n} \propto \sum_{ij} \Big[ \langle n|Q_i|m\rangle \langle m'|Q_j|n'\rangle + \langle n|Q_j|m\rangle \langle m'|Q_i|n'\rangle \\ 
    &-\sum_k \left(\delta_{mn} \langle m'|Q_i|k\rangle \langle k|Q_j|n'\rangle  + \delta_{m'n'} \langle n|Q_i|k\rangle \langle k|Q_j|m\rangle \right)\Big],
\end{align}
where $Q_{i,j}$ are system operators modulated by the interaction with the environment and determining the possible transfer rates. The oscillating terms in Eq. \eqref{eq:masteq} arise from writing the operators $Q_{i,j}$ in interaction picture and have a crucial impact on our conclusions below. Indeed, when integrating Eq. \eqref{eq:masteq}, {\it coherences transferred} from  $\rho_{n'n}$ to $\rho_{m'm}$ vanish on a time-scale $\tau$ of the order of $\hbar/\big(\Delta_{m'n'}-\Delta_{mn}\big)$, due to the rapidly oscillating terms in the exponential. This leads to the common {\it secular approximation}, which in most cases allows one to assume that coherences are not transferred in relaxation processes, unless for pairs of states such that $\Delta_{m'n'}=\Delta_{mn}$. 
Note that in most qudit systems of interest $\tau$ is much shorter than the realistic times required to implement logic operations. For instance, in the molecular electronic spins considered here the anharmonicity is in the 1-50 GHz range, corresponding to $\tau < 1$ ns, while the shortest pulse of the protocol lasts $\gtrsim 10$ ns.  

\begin{figure}[t]
  \centering
  \includegraphics[width=\columnwidth]{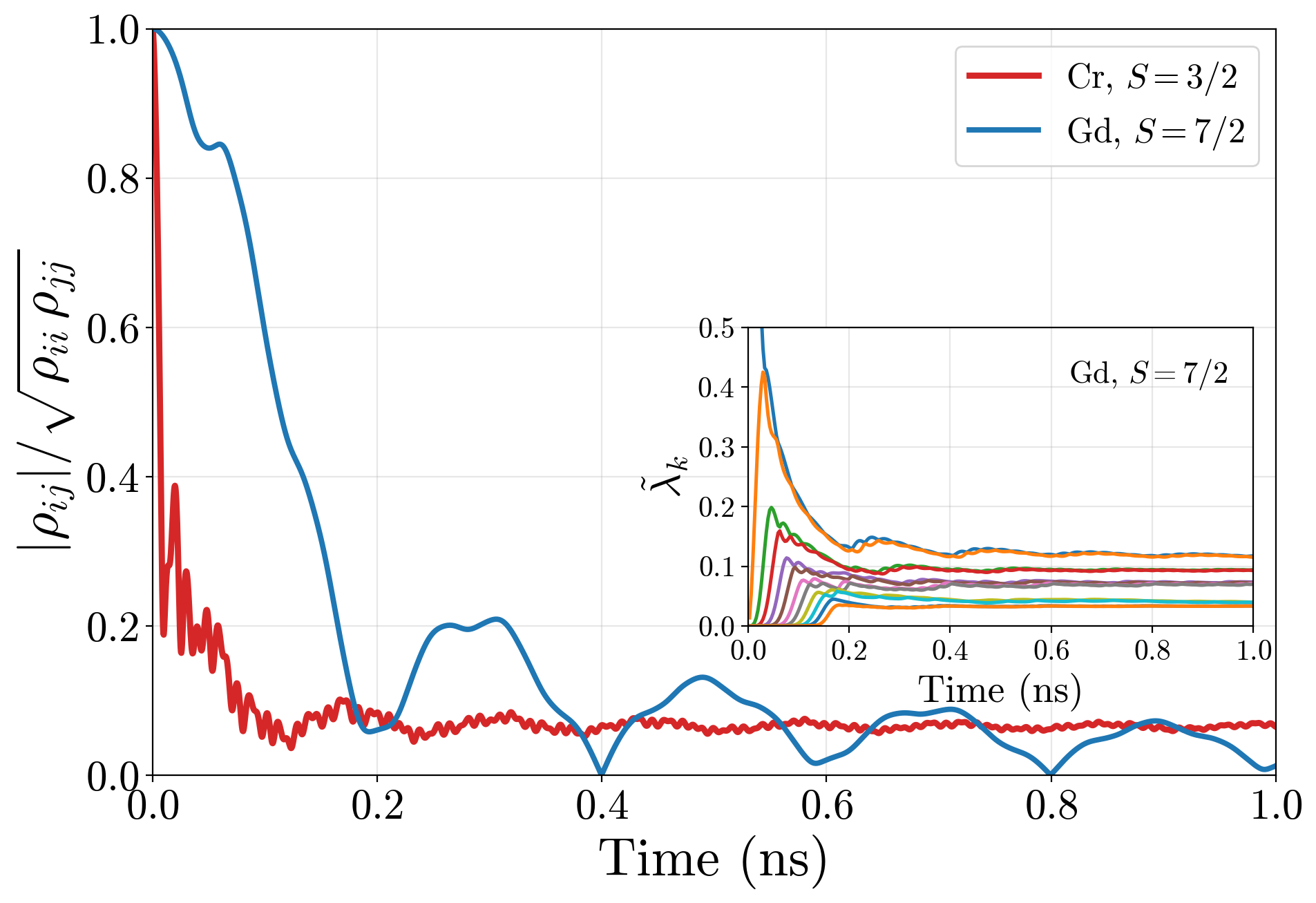}\\
\caption{Relaxation dynamics of electronic spins according to the full
Redfield equation. Main panel: time evolution of the normalised
transferred coherence $|\rho_{ij}|/\sqrt{\rho_{ii}\rho_{jj}}$ on the
initially empty states, for the two systems of
Fig.~\ref{fig:confronto_T1}: the red curve corresponds to Cr,
$D/h=-24.3$ GHz, and the blue one to Gd, $D/h=1.28$ GHz, illustrating the
dependence of the decay on anharmonicity. The transferred coherence is
lost within a fraction of a nanosecond, much shorter than the pulse
durations reported in Fig.~\ref{fig:confronto_T1}. Inset: normalised weights of the Kraus operators,
$\tilde{\lambda}_k = \lambda_k / \sum_{j\neq\mathbb{1}} \lambda_j$ with
$\lambda_k = \mathrm{Tr}(E_k^\dagger E_k)$, obtained by process tomography
of the density-matrix evolution for Gd ($d=8$); the identity component is
excluded from both numerator and sum. As in the main panel, the normalisation removes the overall magnitude of
the decay and isolates its structure, which is set by the anharmonicity
alone. The hierarchy present at the shortest times is lost in less than
$0.2$~ns and the number of relevant errors  exceeds the $d/2=4$ limit of a single-unit code. All simulations
are performed with $T_1=100$ ms.}
\label{fig:relaxation}
\end{figure}

To test how quickly coherences are lost in molecular spins, we initialize some prototypical spin qudits in a superposition of two eigenstates and let them evolve  according to the Redfield equation \eqref{eq:masteq}, i.e. with the inclusion of both relaxation and the coherent Hamiltonian dynamics (introducing phase average). 
As an illustrative example, we consider zero-temperature relaxation driven by the operator \cite{Wurger1998}
\beq
\hat{Q} = \frac{1}{2} \, \big(\hat{O}^1_2 + \hat{O}^2_2 \big) \, ,
\label{SO}
\eeq
with  
$\hat{O}^1_2 =(\hat{S}_x\hat{S}_z+\hat{S}_z\hat{S}_x)/2$ and 
$\hat{O}^2_2=(\hat{S}_x^2 - \hat{S}_y^2)$ Stevens operators of rank 2, yielding 
both $|\Delta m| = 1$ and $|\Delta m| = 2$ transitions. 
Results for normalized transferred coherence in the $S=3/2$ Cr complex considered above \cite{Freedman_Cr} are shown in Fig. \ref{fig:relaxation} (red curve).
The decay is very fast, given the significant anharmonicity, with the
normalized coherence dropping below $0.5$ in approximately $0.008$~ns.
For comparison, the blue curve shows the case of the less anisotropic
GdW$_{10}$ \cite{Jenkins2017}, $D/h = 1.28$~GHz, where the decoherence is significantly slower: the coherence halves in about $0.11$ ns, but still on a time-scale much faster than any practical gate implementation. This separation of timescales, where the coherence decays on a sub-nanosecond scale while the relaxation time is of the order of $T_1 \sim 0.1$ s, ensures that the secular approximation is fully justified.

In this limit we can omit oscillating terms which average to zero and the master equation simplifies to:
\begin{eqnarray}
    \label{eq:MEq_offdiag}
    \dot\rho  = \\ 
     \sum_{n \neq m} k_{m \leftarrow n}\left\{ \ket{m}\bra{n} \rho  \ket{n}\bra{m} -\frac{1}{2} \, \big[\ket{n}\bra{n},\rho \big]_+ \right\} \, .
\end{eqnarray}
Here the first term describes the population growth on the $\ket{m}$ eigenstate, while the second the decay of both coherences and populations associated to $\ket{n}$.
The terms that would analogously lead to coherence growth are canceled by the secular approximation.
For weak, linear coupling to a thermal bosonic bath the transition rates are $k_{m \leftarrow n} \propto | \bra{m} Q \ket{n}|^2 \bar{n}(\Delta_{mn}) $ for $\Delta_{mn} >0 $ (absorption) and $k_{m \leftarrow n} \propto | \bra{m} Q \ket{n}|^2 ( \bar{n}(\Delta_{mn}) + 1)$ for $\Delta_{mn} < 0 $ (emission), where  $\bar{n}(\Delta_{mn})= (e^{ |\Delta_{mn}|/k_B T} - 1 )^{-1}$ is the mean occupation of a bath mode of energy $|\Delta_{mn}|$, and the detailed expression depends also on the bath spectral density.
As the rates are positive, Eq. ~\eqref{eq:MEq_offdiag} is in Lindblad form with jump operators $L_{mn}=\sqrt{k_{m\leftarrow n}}\ket{m}\bra{n}$.

We now focus on the consequences of this coherence loss on QEC. 
Quantum process tomography of a system evolving according to Eq.~\eqref{eq:MEq_offdiag} yields off-diagonal Kraus operators of the form 
\begin{equation}
    \label{eq:Kraus}
    E_{k\equiv\{m,m'\}}  =c_{m,m'} \,\ket{m}\bra{m'},
\end{equation}
i.e. an independent decay channel for each pair of eigenstates $\ket{m},\ket{m'}$ connected by the relaxation operators $Q_{i}$. 
Here we point out a fundamental difference from pure dephasing. Indeed, the master equation \eqref{eq:MEq_diag} for pure dephasing can be recovered by setting $Q_i\equiv S_z$ in the Redfield equation \eqref{eq:masteq} and the resulting Kraus operators preserve a similar diagonal form acting collectively on the whole qudit space. Here, in contrast, the operator $Q$ entering the Redfield equation is split  into several $E_k$ of different form, i.e. {\it the environment  resolves different transitions} \cite{albert2024}.
Crucially, the coefficients $c_{m,m'}$ in Eq. \eqref{eq:Kraus} are comparable to each other, such that no natural hierarchy exists and any unitary transformation would not change significantly their norm \footnote{We could attempt to introduce a hierarchy by changing the Kraus representation of the channel via a unitary transformation \cite{nielsen}. However, one can easily show that the norms of the new operators obtained by unitary transformation starting from a set of orthogonal Kraus operators are between the maximum and minimum norms of the original ones.
Therefore, it is impossible to modify the hierarchy of the Kraus operators by choosing a different representation.
}.  
This lack of a clear hierarchy is illustrated in the inset of Fig. \ref{fig:relaxation}, where process tomography of the Gd system
reveals the emergence, within a fraction of a nanosecond, of several
Kraus operators of the same order of magnitude (one for each independent decay channel). The number of error-operators of the form \eqref{eq:Kraus} of similar norm depends on the specific error, but is typically larger than the maximum number of errors correctable by an embedded encoding ($d/2$). For instance, it is $d-1$ for a simple ladder operator and it can even reach $d(d-1)$ for relaxation operators with all-to-all connectivity. 

These results highlight that relaxation cannot be corrected by an embedded encoding within a single spin, because the system reaches the secular limit too quickly (on a time scale given by the anharmonicity and not by $T_1$) and the resulting number of Kraus operators is too large. We finally note that relaxation-driven errors that lead to operators as in Eq. \eqref{eq:Kraus} can constrain too much the codewords determined by the Knill-Laflamme conditions (see Ref. \cite{albert2024}). 
In fact, the unique solution for them would lead to the exclusion from the code words of any eigenstate subject to relaxation \cite{albert2024}, yielding a decoherence-free subspace. Relatedly, in the bosonic setting, Ref.~\cite[Sec.~VIII.A]{albert2018} showed that a nonlinear Kerr term can strongly alter the QEC performance of codes tailored to pure relaxation.

This poses a fundamental challenge for developing resource-efficient QEC schemes correcting arbitrary (diagonal and non-diagonal) errors and calls for an alternative strategy, not based on embedding QEC within a single spin \footnote{or strongly-interacting multi-spin molecules}.

\subsection{Embedding arbitrary-noise QEC in \\ multi-spin molecules}
We overcome the single-spin
limitations by exploiting the unparalleled chemical tunability of molecular spins.  
This allows us to concatenate several {\it dephasing-tolerant qubits}  within a single molecule, and to apply a second level of encoding based on a multi-qubit code. 
Indeed, once dephasing has been corrected within each $S-\mathcal{s}$ unit, the residual noise acts on each dephasing-tolerant qubit as an effective error, which can be corrected by a stabilizer code \cite{nielsen}.

Compared to a blind approach in which all single-qubit errors are assumed to be equally likely, molecular spins show an important advantage, because the first level of encoding already allows for a huge error reduction, as already evidenced in Fig. \ref{fig:confronto_T1}.
Assuming that the errors due to the residual noise are identical and independent and happen with probability $p$, an ideal decoder that corrects at most $t$ errors, achieves 
\begin{equation}
	P_{err} \sim \binom{n}{t+1}p^{\,t+1} \, .
    \label{eq:PL}
\end{equation}
See SI for a more detailed characterization of the residual noise channel, and for more details on the assumptions behind this error estimate. 
The crucial result is the polynomial amplification: reducing the first-level error from $p$ to a smaller value produces an outer-code error of order $p^{t+1}$.
This scaling is not tied to a particular code family, although the prefactor and practical overhead depend on the specific implementation.
This reduction has immediate practical consequences: a lower number of levels implies smaller codes, reduced overhead, and a more efficient use of the computational resources.
As a result, spin-qudit schemes with embedded error-correction strategies represent a particularly promising platform for fault-tolerant quantum computing.

\begin{figure*}
    \centering
    \resizebox{.9\textwidth}{!}{\input{fig3.tikz}}
    \caption{(a) Scheme of the proposed heptadecanuclear star-shaped molecule  $[\mathrm{W^{V}(CN)_8}\{\mathrm{Ln^{III}}(L)\}_8\{\mathrm{M}(L')\}_8]$ implementing the code presented in the paper. 
    Correction against pure dephasing is embedded at the level of each spin $S_i$ with ancillary electronic spin $\mathcal{s}_i$ (i=1,...,7) while relaxation is addressed by concatenating these units to implement the 7-qubit Steane code into the reported octagonal structure. 
    An octacyanidotungstate(V) core supplies eight cyanido bridges (blue, exchange coupling $J'$) to eight capped lanthanide ions, each of which carries one outer high-spin unit M (red, with coupling $J$).
    Ln is either  $\mathrm{Gd^{III}}$ ($S = 7/2$) or an anisotropic Kramers ion (Yb, Ce, Er) reducing to an effective $\tilde{s} = 1/2$ with $g_\parallel \gg g_\perp$.
    Two variants of the outer unit M are considered: a terminally blocked $\mathrm{Cr^{III}}$ monocyanide, Ln--N$\equiv$C--Cr($L'$), giving an all-cyanide $S = 3/2$ pendant, 
    or a kinetically inert macrocyclic $\mathrm{Gd^{III}}$ chelate linked through a single fluoride bridge, Ln--F--Gd($L'$), giving an $S = 7/2$ pendant.
    Every exchange pathway terminates on a $4f$ center, so both $J$ and $J' \lesssim 1$~cm$^{-1}$; the eight arms share no bridging ligand and lie more than $7$~\AA{} apart, making the interaction negligible. 
    The drawing shows connectivity only: the $[\mathrm{W(CN)_8}]^{3-}$ core is dodecahedral ($D_{2d}$), not planar.
    Ln ions labeled $\mathcal{s}_i$ ($i=1,\dots,7$) are ancillae for syndrome extraction on the dephasing-tolerant qubits, while 
   another ancillary spin $\mathscr{S}_{\mathrm{flag}}$ (gold) is used for the fault-tolerant measurement of stabilizers of the second-level 7-qubit code (yellow), according to circuits like the one shown in panel (b) (the so-called {\it flag} method \cite{Chao2018}).
    The eighth outer high-spin ion M is not used in the protocol.
    (b) Syndrome extraction of a representative weight-4 stabilizer of the $[[7,1,3]]$ code.
    The spins $S_1$--$S_7$ (red) act as logical data qubits; the central ancilla $\mathscr{S}_{\mathrm{syn}}$, prepared in $\ket{+}$ (blue), collects the syndrome; the flag ancilla $\mathscr{S}_{\mathrm{flag}}$, prepared in $\ket{0}$ (gold), detects faults occurring during the extraction.
    The remaining generators are measured by analogous circuits, with the data qubits suitably permuted. } 
    \label{fig:octagon}
\end{figure*}

Because the residual channel contains both diagonal and off-diagonal components of comparable magnitude, we design a molecular setup for handling generic single-qubit errors.
Indeed, we can conceive a molecule as the one sketched in Fig. \ref{fig:octagon}-(a), consisting of 7 spin $\mathcal{s}_i$ ions ($i=1,\dots,7$) and a spin $\mathscr{S}_{\mathrm{flag}}=1/2$, coupled by exchange interactions $J^\prime$ with a central spin $\mathscr{S}_{\mathrm{syn}}=1/2$ in a star-shape structure. Each spin $\mathcal{s}_i$ is coupled by additional exchange interaction $J$ with a spin $S_i$
(the eighth outer high-spin ion M is not used in the protocol).
This system can implement the 7-qubit Steane code, denoted
$[[7,1,3]]$ in the standard $[[n,k_C,d_C]]$ notation for a stabilizer code
encoding $k_C$ logical qubits into $n$ physical qubits with code
distance $d_C$. 
In this case $d_C=3$ allows for the correction of an arbitrary single-qubit error ($t=1$ in Eq.~\eqref{eq:PL}) by exploiting the $S_i-\mathcal{s}_i$ pairs as first-level dephasing-tolerant qubits. 
In particular, each spin $\mathcal{s}_i$ in the external octagon acts as ancilla for stabilization in the code for pure dephasing embedded in each $S_i$.
This is achieved by multi-tone excitations of the spin $\mathcal{s}_i$ conditioned on the {\it error-state} $k$ of the spin $S_i$.
Then, a multi-tone measurement of the electronic spin stabilizes information in the $k-$th error subspace. 
The spin state of each $S_i-\mathcal{s}_i$ pair is factorized by working at a magnetic field sufficiently larger than $J/g_z\mu_B$, thus ensuring  that information encoded in $S_i$ is not corrupted by a {\it non-demolition} measurement of $\mathcal{s}_i$. 

Conversely, the central ancilla $\mathscr{S}_{\mathrm{syn}}$ is used for measuring stabilizers on groups of data qubits according to the {\it flag} protocol of Refs.~\cite{DeAmrit2016,Chao2018,Chamberland2018,Weinstein2015a,Reichardt2021}.
This is achieved by sequences of controlled operations between each data qubit $S_i$ and the ancilla $\mathscr{S}_{\mathrm{syn}}$ prepared in state $\ket{+}$.
The latter is coupled to a second flag ancilla $\mathscr{S}_{\mathrm{flag}}$ to detect errors on the syndrome ancilla that can propagate into correlated errors on the data qubits [see Fig. \ref{fig:octagon}-(b)].
When the flag ancilla is measured in $|1\rangle$, 
additional syndrome measurements determine the appropriate recovery, making fault-tolerant correction of the data errors possible.
Entangling gates between $S_i$ and $\mathscr{S}_{\mathrm{syn}}$ are turned on and off by exploiting the $\mathcal{s}_i$ ancillae as a dynamical switch of the interactions between $S_i$ and $\mathscr{S}_{\mathrm{syn}}$.
Indeed, the energy needed to excite the switch depends on the states of both $S_i$ and $\mathscr{S}_{\mathrm{syn}}$ through the diagonal contributions of the exchange interactions $J$ and $J^\prime$.
Hence, we can implement a conditional excitation of the switch $\mathcal{s}_i$ depending on the state of $S_i$ and $\mathscr{S}_{\mathrm{syn}}$.
This, combined with frequency-selective single-qubit rotations, implements a switchable cNOT gate \cite{Atzori2018,chiesa2024}. \\
The central ancilla $\mathscr{S}_{\mathrm{syn}}$ is then entangled with the flag-qubit $\mathscr{S}_{\mathrm{flag}}$ at one of the vertices of the octagon~\cite{Chao2018,Reichardt2021}.
Finally, both the electronic spin ancillae $\mathscr{S}_{\mathrm{flag}}$ and $\mathscr{S}_{\mathrm{syn}}$ are read out by coupling them with superconducting resonators, as described in \cite{blueprint}.
\begin{figure*}[ht!]
    \centering
    \includegraphics[width=0.9\linewidth]{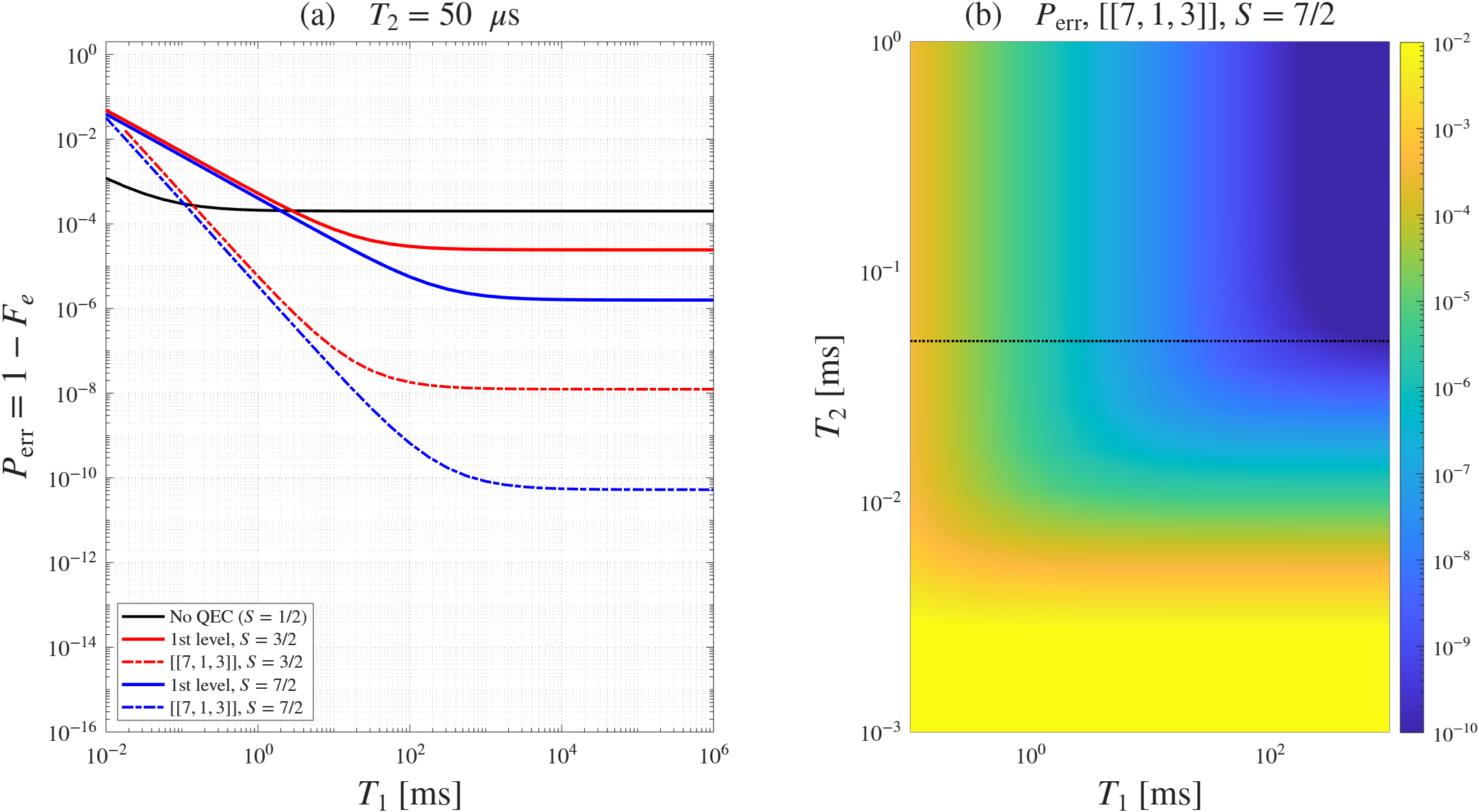}
        \caption{
    Ideal performance of the embedded multi-spin code, with $P_{err}=1-F_e$ the entanglement infidelity of the logical Hadamard gate.
     (a) $P_{\rm err}$ at fixed $T_2 = 50\,\mu$s, as a function of $T_1$.
The black curve is an uncorrected two-level system (no QEC); solid and
dashed curves correspond to first-level dephasing
correction only and second-level $[[7,1,3]]$ concatenation, with red curves for the
$S=3/2$ Cr$^{\rm III}$  and blue curves for the $S=7/2$ Gd$^{\rm III}$. (b) $P_{\rm err}$ for the $[[7,1,3]]$ code on $S=7/2$  as a
function of the relaxation time $T_1$ and the dephasing time $T_2$; the
dotted black line marks the cut shown in panel (a). }
    \label{fig:seven}
\end{figure*}

The pattern of required interactions is sketched in Fig. \ref{fig:octagon}-(a), with $S_i=5/2$ or $7/2$, $\mathcal{s}_i\ge1/2$ and small exchange interactions between all spin pairs, significantly weaker than their difference in Zeeman energy. This is important to ensure factorization both of the $S_i-\mathcal{s}_i$ and of the $S_i-\mathscr{S}_{\mathrm{syn}}$ states in a sizeable magnetic field and hence effectively suppress residual interactions when the switch is off \cite{chiesa2024}.
The use of spins $\mathcal{s}_i=1/2$ as ancillae for measuring stabilizers of the embedded code (instead of higher spin ions as in \cite{mezzadri2024}) makes the correction circuits longer but does not compromise fault tolerance. \\
An important requirement is that the individual units to be combined in the 7-qubit code should interact only with $\mathscr{S}_{\mathrm{syn}}$ and not directly between each other.

Before moving to a concrete proposal for a molecule, we mention that to achieve universal FT quantum computing with the 7-qubit Steane code we adopt, at least one of the fundamental gates must be implemented via general protocols known as magic-state preparation~\cite{Goto2016a,Chamberland2019} and magic state distillation~\cite{Bravyi2005,Bravyi2012}.
While this approach is standard, it often incurs a high overhead, and engineering also these steps to fully take advantage of the high connectivity of the proposed molecular architecture is left for future work.

{\bf Proposed molecule.} The architecture we propose is a heptadecanuclear star, $[\mathrm{W^{V}(CN)_8}\{\mathrm{Ln^{III}}(L)\}_8\{\mathrm{M}(L')\}_8]$, in which a single octacyanidotungstate(V) core provides the central spin, eight lanthanide ions occupy the arms, and one high-spin unit M caps each arm.
The choice of an eight-coordinate cyanidometalate is dictated by the required connectivity:
$[\mathrm{M^{V}(CN)_8}]^{3-}$ (M $=$ Mo, W) is the only routinely accessible molecular node presenting eight potential bridges, and the simultaneous
$\mu$-coordination of all eight cyanido ligands is documented in the
three-dimensional frameworks
$[\mathrm{M^{IV}}\{(\mu\text{-CN})_4\mathrm{Mn^{II}(H_2O)_2}\}_2]_n$
(M $=$ Nb, Mo, W)~\cite{octacyano3D}. We stress that this is an ambitious target and smaller stars (e.g. hexagonal) would meet many of the requirements for the QEC protocol outlined; our intention is to provide synthetic chemists with a new synthetic target with the potential to significantly advance QEC.
The proposal is based on previous heptanuclear stars, initially reported by Mallah and co-workers \cite{Mallah1995,Scuiller1996, Rogez2000}, extended by several other groups \cite{Parker2001, marvaud2003, Glaser2006}. The strategy involves a central $[\mathrm{Cr}^{\mathrm{III}}(\mathrm{CN})_6]^{3-}$ unit which uses the cyanides to bind to peripheral unit which are terminally saturated by using a pentadentate ligand giving $[\mathrm{Cr}\{\mathrm{M}^{\mathrm{II}}(\mathrm{tetren})\}_6]^{9+}$
 \cite{Mallah1995}. The six peripheral ions are mutually non-interacting \cite{Mallah1995}. Here we transpose this to $[\mathrm{W}^{\mathrm{V}}(\mathrm{CN})_8]^{3-}$ to produce an eight-pointed star. We propose attaching
$\mathrm{Ln}^{\mathrm{III}}$ ions encapsulated in a hexa- or
heptadentate ligand that leaves exactly two coordination
positions vacant in trans positions; such
$\mathrm{Ln}^{\mathrm{III}}$ compounds have been reported
using macrocyclic ligands \cite{Song2022,Canaj2019}.
The Ln--N$\equiv$C--W$^{\mathrm{V}}$ linkage was established in 
$\{\mathrm{Ln^{III}W^{V}}\}$ binuclear complexes~\cite{alexandru2017} and in
polynuclear Ln--W(CN)$_8^{3-}$ assemblies~\cite{lnw2007}. While molecular stars based on
$[\mathrm{W}^{\mathrm{V}}(\mathrm{CN})_8]^{3-}$
are not known, this unit is used in discrete high-spin
molecules \cite{Zhong2000,Larionova2000}, and stars based
on related heptacyanide building blocks are known
\cite{Freedman2008}. Another possibility would be to
replace $\mathrm{Ln}^{\mathrm{III}}$ in the middle of each arm
with a weakly coupled $3d$ ion such as
$\mathrm{Fe}^{\mathrm{III}}$ or $\mathrm{Mn}^{\mathrm{II}}$;
an example of the required unit has been reported
\cite{Zhang2009}. The only essential requirement for the
QEC protocol is a small exchange coupling. In the final step, each $\mathrm{Ln}^{\mathrm{III}}$ ion is
connected to an outer metal unit ($M$ in Fig.\ref{fig:octagon}).
Two realisations of the outer unit are available.
In variant \textbf{(a)} M $=$ $\mathrm{Cr^{III}}$ and is delivered as a terminally
blocked monocyanide, so that the whole molecule is cyanido-bridged throughout;
the $3d^3$ configuration of
$\mathrm{Cr^{III}}$ supplies both inertness due to a high crystal field stabilisation energy and the highest spin 
($S = 3/2$) accessible from a cyanidometalate [the strong-field cyanide
rendering $\mathrm{Fe^{III}}$ and $\mathrm{Mn^{II}}$ low-spin].
This linkage has direct precedent in cyanido-bridged
$\mathrm{Gd^{III}}$--$\mathrm{Cr^{III}}$ molecules~\cite{gdcr2002}.
In variant \textbf{(b)} M $=$ $\mathrm{Gd^{III}}$ is introduced as a preformed
macrocyclic chelate of the DOTA/DO3A family and attached through a single
fluorido bridge; the well-known kinetic inertness of such
chelates~\cite{caravan1999} allows use of a lanthanide as M as well as Ln (Figure \ref{fig:octagon}); the anionic polyaminocarboxylate caps keep the overall charge of the cluster moderate, in contrast to the highly cationic all-cyanide assembly. In both cases the competing outcome is polymerisation ~\cite{perrier2012} which the polydentate ligand is designed to suppress.\\
The central $5d^1$ $\mathrm{W^{V}}$ ion ($S_{\rm syn}$ in Figure \ref{fig:octagon}) is an isotropic $S = 1/2$ centre
($g \approx 1.97$). Two choices are available for the $\mathrm{Ln^{III}}$ sites: with
$\mathrm{Ln^{III}}=\mathrm{Gd^{III}}$ the molecule is a nearly isotropic Heisenberg star of eight
$S = 7/2$ arms, whereas with an anisotropic Kramers ion (Yb, Ce, Er) each arm
reduces, at the temperatures of interest, to a well-isolated axial doublet
described by an effective $\tilde{s} = 1/2$ with
$g_\parallel \gg g_\perp$. The outer M sites carry $S = 3/2$ in variant~(a) and
$S = 7/2$ in variant~(b).
In both cases, because every exchange pathway terminates on a $4f$ centre, the
contracted $4f$ shell keeps both the radial coupling $J'$
(W--C$\equiv$N--Ln) and the outer coupling $J$ below $1$~cm$^{-1}$ (Figure \ref{fig:octagon}); the two are
not equal, the cyanido-bridged Ln--$\mathrm{Cr^{III}}$ pathway of variant~(a)
being the stronger and the single fluorido bridge of variant~(b) the weaker,
since Ln--Ln exchange never exceeds $\sim$$0.1$~cm$^{-1}$ even across a linear
Ln--F--Ln unit~\cite{corredoira2022}.
By contrast the arms share no bridging ligand and the terminal M sites will be more than  $7$~\AA{} apart, reducing interactions between the M-sites to a dipolar residue of order  $10^{-2}$~cm$^{-1}$, and
the exchange topology is a genuine spin star.
In the anisotropic case, projection of the exchange onto the axial
doublets~\cite{chibotaru2015} renders both couplings Ising-like along the local
easy axes, which are not collinear: since the $[\mathrm{W(CN)_8}]^{3-}$ core is dodecahedral
($D_{2d}$) rather than $C_8$-symmetric, the eight cyanido positions split into
two inequivalent sets of four, and in a fixed applied field each arm acquires a
distinct effective splitting,
making the arms individually addressable by frequency.

\textbf{Performance analysis.} We now compute the ideal performance of the proposed embedded protocol in correcting both dephasing and relaxation.
The logical error probability follows the same polynomial
law of Eq.~\eqref{eq:PL} and results
are shown in Fig.~\ref{fig:seven}.
Panel (a) shows $P_\mathrm{err}$ as a function of $T_1$ at the fixed value $T_2=50\,\mu\mathrm{s}$, for the uncorrected qubit, for
the first-level dephasing correction alone, and for its concatenation 
with the $[[7,1,3]]$ code, on $S=3/2$ (Cr) and $S=7/2$ (Gd). We remark that this value of $T_2$ (already measured in V and Cu complexes after proper chemical optimization \cite{BaderNatComm14,atzoritesi,Zadrozny}) is realistic for isolated molecules, given the noteworthy increase with dilution recently discussed in \cite{Conor2025}.

All corrected curves share the same qualitative behavior, saturating at large $T_1$ to a
$T_1$-independent plateau set by the residual dephasing error left
uncorrected by the first-level embedded code once relaxation becomes
negligible; this plateau is reached for $T_1$
of the order of seconds.
The plateau level, however, differs by several
orders of magnitude between the different curves, reflecting the
progressively stronger suppression achieved by larger spin.\\
At $T_2 = 50 \mu s$, $S=7/2$ outperforms
$S=3/2$ at the first level, since the larger spin allows
a more effective embedded suppression of the leading dephasing
channel; this initial gap is then amplified at the second level. 
Panel (b) shows that, for $S=7/2$, the qualitative behavior of panel (a) holds across the entire $(T_1, T_2)$ plane.  
The advantage is lost only when the correction cycle takes long
enough, relative to $T_1$ and $T_2$, for its own residual error to
compete with the suppression achieved by the code.\\
This confirms on the complete $(T_1,T_2)$ landscape that the
hybrid embedded/concatenated strategy remains advantageous across the whole range of coherence times relevant for present-day molecular
spins.

\section{Conclusions}
We have introduced an embedded scheme for FT quantum computing against generic (diagonal and non-diagonal) noise, exploiting multi-spin molecules as building blocks. 
 
Having highlighted the limitations of anharmonic units (such as effective spins $S$ with zero-field splitting) in embedding correction of non-diagonal errors at the single-spin level, we have shown that quantum error correction can be embedded within single objects by exploiting multi-spin molecules for a two-level encoding: in the first level spins $S\ge 3/2$ are used to correct leading pure dephasing errors with the help of exchange-coupled spin ancillae. These dephasing-tolerant qubits are then combined into a multi-spin molecule where a multi-qubit code is used to handle residual non-diagonal errors. This approach exploits the unparalleled  possibility offered by molecular spins to be arranged into tailored multi-spin clusters, which translates into the unique capability of embedding quantum logical units protected against generic noise. 
Note that the remarkable correction at the first level of encoding allows us to achieve an impressive error suppression with limited overhead, i.e. even with a limited number of spins in the second level of encoding. Hence, we have proposed a hypothetical star-shaped molecule with the proper connectivity for implementing the proposed protocol. \\
Other schemes could be conceived, both at the synthetic and algorithmic level. For instance, we could consider codes tailored to the form of the residual noise, which is similar to generalized amplitude damping~\cite{Cafaro2014} or implementations based on the 9-qubit Bacon-Shor code or on the more compact 5-qubit code. The latter could simplify the chemical synthesis, at the price of more complex manipulation sequences for logical gates. This is relevant because heptanuclear stars with a single paramagnetic core surrounded by a shell of six ions are already available, providing a natural starting point for the required star-like connectivity. In particular, five of the external sites could in principle host the physical qubits of the $[[5,1,3]]$ code. \\
Further error suppression could be achieved by adding dephasing-tolerant qubits at the second level of encoding, e.g. by replacing the [[7,1,3]] with the higher-distance $[[11,1,5]]$ code. \\
The estimated performance relies on several assumptions on the gate implementation, which in turn depend on the capability of designing suitable molecules. 
For instance, imperfections in gate or measurement implementations at the single qubit level are expected to limit the performance of the outer code, but not to compromise its fault-tolerance. Conversely, 
small interactions between edge qubits or residual couplings after the implementation of two-qubit gates (needed for the stabilizer measurements) could lead to correlated errors not managed by the 7-qubit code.

Extending this approach beyond a single second-level logical qubit requires entangling gates between distinct molecules.
The chemically assembled logical units we have proposed could be coupled via superconducting resonators~\cite{blueprint}, thus providing switchable interactions for logical CZ gates. 
For the Steane code, these gates can be implemented through pairwise logical CZ operations between corresponding dephasing-tolerant first-level qubits.
However, 
developing a full-fledged fault-tolerant implementation of these gates is left for future work.

The high correction capability of the proposed scheme relies also on a striking hierarchy between the different sources of noise, with pure dephasing largely dominant and corrected at the embedded single-spin level.
This two-level concatenation strategy also offers an interesting route toward quantum error correction in atomic platforms.
In particular, neutral atoms with a large nuclear spin, such as $^{87}\mathrm{Sr}$, provide a natural setting for an embedded qudit encoding.
Atomic control techniques such as tensor light shifts and Rydberg-mediated entangling operations~\cite{deutsch2024} are the building blocks for such an extension.
Adapting our first-level dephasing-correction protocol requires engineering an anharmonicity with suitable connectivity for logical operations and syndrome extraction.
However, photon scattering and errors associated with the additional control fields need to be included.
The induced anharmonicity will generally also allow the environment to resolve (partially or fully) distinct dissipative transitions into independent errors. 
The resulting residual logical errors could be addressed by the concatenation proposed here, provided they remain sufficiently weak and predominantly local.
Moreover, the high degree of individual addressability in atomic arrays could also facilitate alternative implementations of the second-level encoding.
We notice that this would go beyond the large spin-cat approach of Ref.~\cite{deutsch2024}, which assumes the dominant physical errors act as a single spin operator on the whole qudit space.
Overall, these considerations motivate further work toward atomic implementations of the proposed scheme. A detailed microscopic assessment of the driven dynamics and the resulting logical noise is the next step toward conclusively establishing the feasibility of this approach.\\

\emph{Note} -- 
during the submission of the present
manuscript, we got aware of another work \cite{albert2026} on closely related topics
and appeared few days before.

{\hfill\\ \bf Acknowledgments --} 
L.L. is pleased to thank Francesco Minardi, Silvia Neri, and Andrea Trombettoni for fruitful discussions. 
The work received funding from the European Union—NextGenerationEU, PNRR MUR Project PE0000023-NQSTI, 
F.A., A.C. and S.C. acknowledge support from the Project PN-RIC PIMIQ (MI1741, B42F26000460005), 
M.M. acknowledges funding from the European Union—NextGenerationEU under the National Recovery and Resilience Plan (NRRP), Mission 4 Component 1 Investment 3.4 and 4.1. Decree by the Italian Ministry N. 351/2022 CUP
D92B22000530005.
R.E.P.W. thanks EPSRC(UK) for the responsive mode grant EP/Z535801/1.

\clearpage
\onecolumngrid
\begin{center}
{\large\textbf{Supplementary Information for\\[0.2em]
Beyond Pure Dephasing: Quantum Error Correction in\\
Single Molecules Requires Multiple Spins}}\\[1.2em]
\end{center}
\vspace{1em}

\setcounter{section}{0}
\setcounter{figure}{0}
\setcounter{equation}{0}
\renewcommand{\thesection}{S\arabic{section}}
\renewcommand{\thefigure}{S\arabic{figure}}
\renewcommand{\theequation}{S\arabic{equation}}
   \renewcommand{\theHfigure}{S\arabic{figure}}
\section{Entanglement fidelity and logical error}
\label{app:entanglement_fidelity}

In this paper we quantify the accuracy of a logical operation by its entanglement fidelity $F_e$ with respect to the ideal gate. The noisy gate is described by a quantum channel $\mathcal{E}_L$ on the logical qubit, which includes the noisy operation and error recovery, with measurement outcomes averaged after the appropriate corrections. For a target ideal unitary $U_L$ it is useful to define the error channel
\begin{equation}
    \Lambda(\rho)=U_L^\dagger\mathcal{E}_L(\rho)U_L \, ;
\end{equation}
for an ideal implementation this would be the identity channel $\mathcal{I}$. When no gate is performed, e.g. for a quantum memory, one can set $U_L=\openone$. The qubit here is the decoded logical degree of freedom, regardless of the dimension of the molecular spins used to encode it.

The entanglement fidelity is defined as~\cite{nielsenSi}:
\begin{equation}
    F_e=\bra{\Phi}(\mathcal{I}_R\otimes\Lambda)
        (\ket{\Phi}\bra{\Phi})\ket{\Phi} \, ,
    \label{eq:appendix_entanglement_fidelity}
\end{equation}
where $R$ is a reference qubit, and $\ket{\Phi}=\frac{\ket{0}_R\ket{0_L}+\ket{1}_R\ket{1_L}}{\sqrt{2}}$ is the maximally entangled state. This quantity measures how well the operation preserves quantum information, including correlations with another system, rather than testing only one chosen input state. We stress that the reference qubit is a mathematical construction in the definition and is not physically related to the molecular architecture. By writing the Kraus representation of the channel $\mathcal{E}_L(\rho)=\sum_a A_a\rho A_a^\dagger$, the same quantity is expressed as
\begin{equation}
    F_e=\frac{1}{4}\sum_a
        \left|\operatorname{Tr}(U_L^\dagger A_a)\right|^2,
    \label{eq:appendix_fidelity_kraus}
\end{equation}
which follows from $\bra{\Phi}(\openone\otimes B)\ket{\Phi}=\operatorname{Tr}(B)/2$.

In the main text, we use the entanglement infidelity $P_{\rm err}=1-F_e$ as the logical error measure. This interpretation is exact for a Pauli error channel, defined as
\begin{equation}
    \Lambda(\rho)=p_I\rho+p_X X_L\rho X_L
        +p_Y Y_L\rho Y_L+p_Z Z_L\rho Z_L,
\end{equation}
where $p_I + p_X + p_Y + p_Z = 1$. Since each non-identity Pauli operator maps $\ket{\Phi}$ to a state orthogonal to it, it follows that
\begin{equation}
    F_e=p_I,\qquad P_{\rm err}=p_X+p_Y+p_Z.
    \label{eq:appendix_pauli_error}
\end{equation}
For this class of channels $P_{\rm err}$ is exactly the probability of a non-identity logical Pauli error. For a general channel, entanglement infidelity remains a useful error measure but need not have this statistical interpretation. Replacing the error channel by its Pauli twirl preserves $F_e$, and therefore gives a Pauli model with the same $P_{\rm err}$.

The entanglement fidelity also has an alternative interpretation. By averaging the fidelity with the ideal output state uniformly over input states distributed on the Bloch sphere, we obtain the average fidelity, which has a simple relation to the entanglement fidelity~\cite{Nielsen2002AverageFidelity}:
\begin{equation}
    \overline{F}=\int d\psi\,\bra{\psi}\Lambda(\ket{\psi}\bra{\psi})\ket{\psi},
    \qquad \overline{F}=\frac{2F_e+1}{3}.
    \label{eq:appendix_average_entanglement_relation}
\end{equation}
Interestingly, the uniform average over the Bloch sphere can be evaluated by taking a uniform average over the six eigenstates of the logical Pauli operators~\cite{Bowdrey2002Fidelity}:
\[
\left\{\ket{0_L},\ket{1_L},
\frac{\ket{0_L}\pm\ket{1_L}}{\sqrt{2}},
\frac{\ket{0_L}\pm i\ket{1_L}}{\sqrt{2}}\right\}.
\]

\section{Effective noise channel on the first-level dephasing-tolerant qubits}
\label{app:Effective-noise-channel}

\begin{figure}[tbp]
    \centering
    \includegraphics[width=0.5\linewidth]{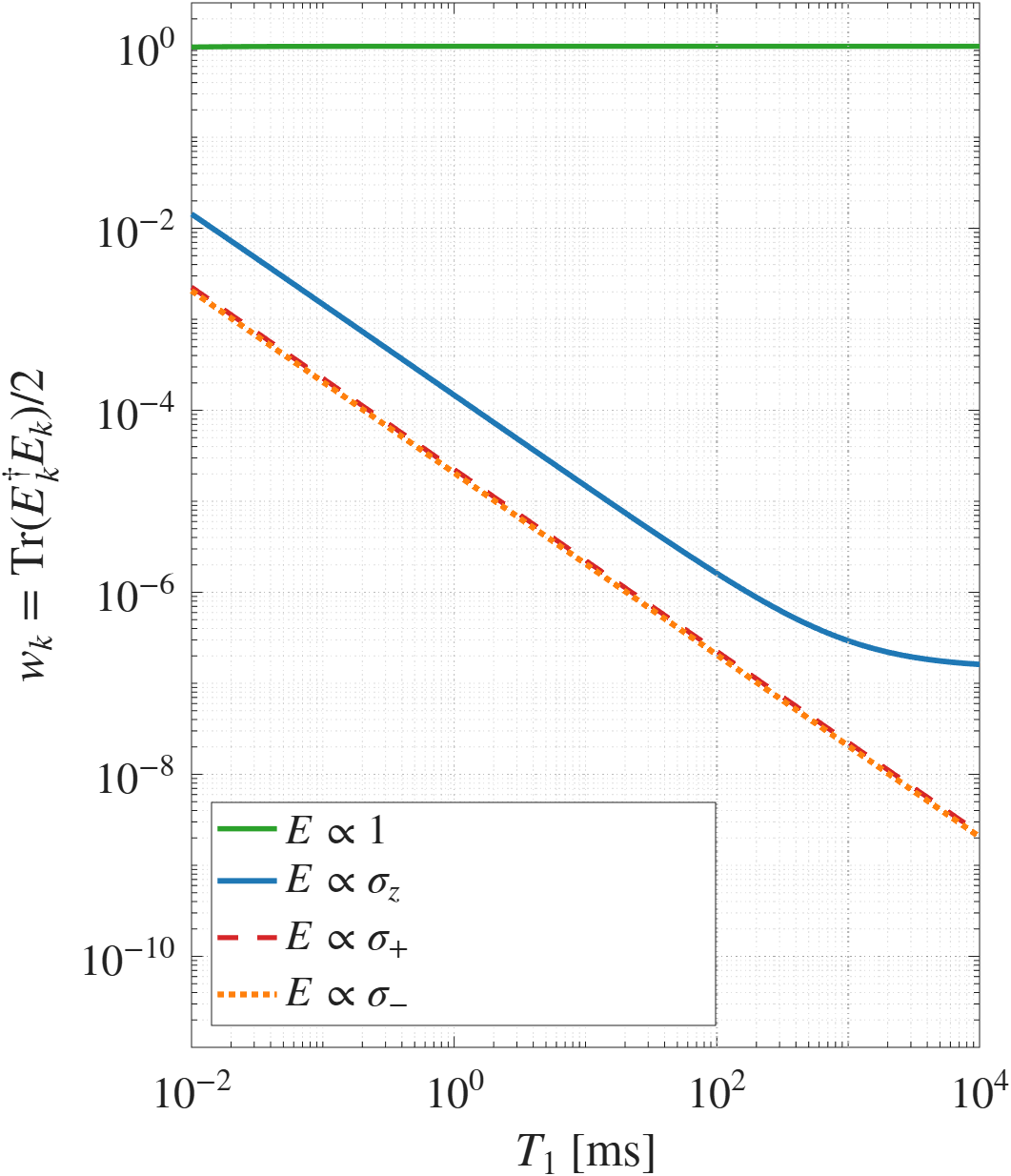}
    \caption{Structure of the residual noise acting on a first-level logical qubit. Weights $w_k = \mathrm{Tr}(E_k^\dagger E_k)/2$ of the Kraus operators of the logical channel, as a function of the relaxation time $T_1$ at fixed $T_2 = 50\,\mu$s, for the $S=7/2$ Gd(III) dephasing-tolerant unit of Fig.~\ref{fig:confronto_T1} of the main text. The channel is the one generated by the full correction cycle of Fig.~\ref{fig:confronto_T1} of the main text for a logical qubit subjected to one correction cycle with no logical rotation applied, with the codewords designed against pure dephasing alone and kept fixed as $T_1$ is varied. The four canonical Kraus operators are found to be proportional to $\mathbb{1}$, $\sigma_z$, $\sigma_+$ and $\sigma_-$. The weights of the $\sigma_+$ and $\sigma_-$ components coincide to within $9\%$ and scale as $1/T_1$, whereas the $\sigma_z$ component saturates at large $T_1$ to the residual dephasing error left uncorrected by the embedded code.}
    \label{fig:canale}
\end{figure}

Process tomography of the residual noise channel acting on the dephasing-tolerant units shows a phase-covariant structure, with two diagonal Kraus components in the span of $\openone$ and $\sigma_z$ and transition components proportional to $\sigma_+$ and $\sigma_-$. In Fig.~\ref{fig:canale} we show the weights of the Kraus operators as a function of $T_1$ for a fixed value of $T_2$.

To obtain an estimate of the error for further concatenation, we replace the channel by its Pauli twirl,
\[
    \mathcal{E}_{\mathrm{P}}(\rho)=\frac{1}{4}
    \sum_{i=0}^{3}\sigma_i\mathcal{E}(\sigma_i\rho\sigma_i)\sigma_i,
    \qquad \sigma_0=\openone,
\]
to obtain a Pauli channel. The twirling operation preserves the entanglement fidelity and is exact when the residual logical noise is Pauli. For identical and independent twirled errors with total probability $p=p_X+p_Y+p_Z \ll 1$, an ideal decoder of distance $d_C=2t+1$ that corrects every pattern of weight at most $t$ gives:
\begin{equation}
    P_{err}=\sum_{j=t+1}^{n}\binom{n}{j}p^j(1-p)^{n-j}
    \sim\binom{n}{t+1}p^{\,t+1} \, .
    \label{eq:PLa}
\end{equation}
Numerical evidence shows that, in the parameter regime considered in this work, the Pauli twirling approximation captures the behaviour of the error probability well.

\end{document}

%% file: fig3.tikz
%
%
%
\begingroup%
\providecommand{\FigFourShowMolecule}{1}%
\providecommand{\FigFourShowCircuit}{1}%
\providecommand{\FigFourShowLegend}{1}%
\providecommand{\FigFourMoleculeLabel}{(a)}%
\providecommand{\FigFourCircuitLabel}{(b)}%
\providecommand{\FigFourOmitBottomM}{0}%
\providecommand{\FigFourCircuitDataColour}{f4data}%
%
\def\fFourLnR{1.80}%
\def\fFourOuterR{3.28}%
\def\fFourWR{0.43}%
\def\fFourLnNodeR{0.325}%
\def\fFourMNodeR{0.30}%
\def\fFourOuterLabelR{3.94}%
\def\fFourInnerLabelOut{0.56}%
\def\fFourInnerLabelTan{0.34}%
%
\def\fFourCircuitTop{4.00}%
\def\fFourRowPitch{0.45}%
\def\fFourGateFirst{0.78}%
\def\fFourGatePitch{0.88}%
\def\fFourHColumn{5.66}%
\def\fFourMeasureColumn{6.34}%
\def\fFourWireEnd{6.66}%
\def\fFourTargetRadius{0.160}%
\def\fFourControlRadius{0.056}%
\def\fFourCircuitShift{6.05}%
\def\fFourLegendLeft{-1.35}
\def\fFourLegendTop{-0.10}
%
\definecolor{f4ink}{HTML}{252B31}%
\definecolor{f4blue}{HTML}{285D80}%
\definecolor{f4blueedge}{HTML}{163D57}%
\definecolor{f4gold}{HTML}{A77900}%
\definecolor{f4goldedge}{HTML}{745719}%
\definecolor{f4outerbond}{HTML}{555C66}%
\definecolor{f4rule}{HTML}{D7DCE2}%
\definecolor{f4subtle}{HTML}{5A6470}%
\definecolor{f4outer}{HTML}{C73544}%
\definecolor{f4outeredge}{HTML}{8F2530}%
\colorlet{f4data}{f4outer}%
\colorlet{f4switchtext}{f4gold}%
\begin{tikzpicture}[
  x=1cm,y=1cm,line cap=round,line join=round,
  every node/.style={text=f4ink,inner sep=0pt,outer sep=0pt},
  f4/core bond/.style={draw=f4blue,line width=1.20pt},
  f4/outer bond/.style={draw=f4outerbond,line width=1.00pt},
  f4/data label/.style={text=f4data,font=\normalfont\fontsize{12}{14}\selectfont},
  f4/switch label/.style={text=f4switchtext,font=\normalfont\fontsize{12}{14}\selectfont},
  f4/ancilla label/.style={text=f4blue,font=\normalfont\fontsize{12}{14}\selectfont},
  f4/flag label/.style={f4/ancilla label,text=f4gold},
  f4/panel label/.style={anchor=north west,font=\normalfont\fontsize{11}{13}\selectfont},
  f4/data wire/.style={draw=\FigFourCircuitDataColour,line width=0.82pt},
  f4/ancilla wire/.style={draw=f4blue,line width=0.92pt},
  f4/flag wire/.style={f4/ancilla wire,draw=f4gold},
  f4/gate/.style={draw=f4ink,line width=0.82pt},
  f4/key/.style={anchor=base west,align=left,font=\normalfont\fontsize{8.6}{10.5}\selectfont},
  f4/subkey/.style={anchor=base west,align=left,text=f4subtle,font=\normalfont\fontsize{8.0}{10.0}\selectfont},
  f4/variant/.style={anchor=base west,font=\normalfont\fontsize{8.2}{10.2}\selectfont},
  pics/f4 outer/.style={code={
    \filldraw[fill=f4outer,draw=f4outeredge,line width=0.65pt]
      (0,0) circle[radius=\fFourMNodeR];
    \node[text=white,font=\normalfont\fontsize{10}{12}\selectfont]
      at (0,0) {$\mathrm{M}$};
  }},
  pics/f4 target/.style={code={
    \draw[f4/gate,fill=white] (0,0) circle[radius=\fFourTargetRadius];
    \draw[f4/gate] (-\fFourTargetRadius,0) -- (\fFourTargetRadius,0);
    \draw[f4/gate] (0,-\fFourTargetRadius) -- (0,\fFourTargetRadius);
  }},
  pics/f4 hadamard/.style={code={
    \draw[f4/gate,fill=white] (-0.22,-0.22) rectangle (0.22,0.22);
    \node[font=\normalfont\fontsize{11.5}{13}\selectfont] at (0,0) {$\mathrm{H}$};
  }},
  pics/f4 meter/.style={code={
    \draw[f4/gate,fill=white] (-0.28,-0.20) rectangle (0.28,0.20);
    \draw[f4/gate,line width=0.70pt] (-0.19,-0.085) arc[start angle=170,end angle=10,radius=0.195];
    \draw[f4/gate,line width=0.70pt,->,>=stealth] (-0.03,-0.13) -- (0.13,0.14);
  }}
]

\path[use as bounding box] (-4.35,-3.70) rectangle (14.10,4.30);

\ifnum\FigFourShowMolecule=1\relax
  \ifx\FigFourMoleculeLabel\empty\else
    \node[f4/panel label] at (-4.25,4.18) {\FigFourMoleculeLabel};
  \fi
  \foreach \i in {0,...,7} {
    \pgfmathsetmacro{\fFourAngle}{90-45*\i}
    \coordinate (f4-ln-\i) at (\fFourAngle:\fFourLnR);
    \coordinate (f4-m-\i) at (\fFourAngle:\fFourOuterR);
    \draw[f4/core bond] (0,0) -- (f4-ln-\i);
    \ifnum\i=4\relax
      \ifnum\FigFourOmitBottomM=0\relax
        \draw[f4/outer bond] (f4-ln-\i) -- (f4-m-\i);
      \fi
    \else
      \draw[f4/outer bond] (f4-ln-\i) -- (f4-m-\i);
    \fi
  }
  \foreach \i in {0,...,7} {
    \filldraw[fill=f4gold,draw=f4goldedge,line width=0.70pt] (f4-ln-\i) circle[radius=\fFourLnNodeR];
    \node[text=white,font=\normalfont\fontsize{10}{12}\selectfont] at (f4-ln-\i) {$\mathrm{Ln}$};
    \ifnum\i=4\relax
      \ifnum\FigFourOmitBottomM=0\relax
        \pic at (f4-m-\i) {f4 outer};
      \fi
    \else
      \pic at (f4-m-\i) {f4 outer};
    \fi
  }
  \filldraw[fill=f4blue,draw=f4blueedge,line width=0.80pt] (0,0) circle[radius=\fFourWR];
  \node[text=white,font=\normalfont\fontsize{12}{14}\selectfont] at (0,0) {$\mathrm{W}$};
  \foreach \i/\k in {0/4,1/5,2/6,3/7,5/1,6/2,7/3} {
    \pgfmathsetmacro{\fFourAngle}{90-45*\i}
    \pgfmathsetmacro{\fFourLabelDX}{\fFourInnerLabelOut*cos(\fFourAngle)+\fFourInnerLabelTan*sin(\fFourAngle)}
    \pgfmathsetmacro{\fFourLabelDY}{\fFourInnerLabelOut*sin(\fFourAngle)-\fFourInnerLabelTan*cos(\fFourAngle)}
    \node[f4/switch label] at ([xshift=\fFourLabelDX cm,yshift=\fFourLabelDY cm]f4-ln-\i) {$\mathcal{s}_{\k}$};
    \node[f4/data label] at (\fFourAngle:\fFourOuterLabelR) {$S_{\k}$};
  }
  \node[f4/flag label,anchor=east] at ([xshift=-0.14cm,yshift=-0.42cm]f4-ln-4) {$\mathscr{S}_{\mathrm{flag}}$};
  \node[f4/ancilla label,anchor=east] at (-0.45,0.22) {$\mathscr{S}_{\mathrm{syn}}$};
  \node[text=f4blue,font=\normalfont\fontsize{12}{14}\selectfont] at (0.93,0.40) {$J'$};
  \node[text=f4outerbond,font=\normalfont\fontsize{12}{14}\selectfont] at (1.65,2.10) {$J$};
\fi

\ifnum\FigFourShowCircuit=1\relax
  \begin{scope}[xshift=\fFourCircuitShift cm]
    \ifx\FigFourCircuitLabel\empty\else
      \node[f4/panel label] at (-1.35,4.18) {\FigFourCircuitLabel};
    \fi
    \foreach \k in {1,...,7} {
      \pgfmathsetmacro{\fFourY}{\fFourCircuitTop-(\k-1)*\fFourRowPitch}
      \draw[f4/data wire] (0,\fFourY) -- (\fFourWireEnd,\fFourY);
      \node[f4/data label,anchor=east] at (-0.14,\fFourY) {$S_{\k}$};
    }
    \pgfmathsetmacro{\fFourSyndromeY}{\fFourCircuitTop-7*\fFourRowPitch}
    \pgfmathsetmacro{\fFourFlagY}{\fFourCircuitTop-8*\fFourRowPitch}
    \draw[f4/ancilla wire] (0,\fFourSyndromeY) -- (\fFourWireEnd,\fFourSyndromeY);
    \draw[f4/flag wire] (0,\fFourFlagY) -- (\fFourWireEnd,\fFourFlagY);
    \node[f4/ancilla label,anchor=east] at (-0.14,\fFourSyndromeY) {$\mathscr{S}_{\mathrm{syn}}:\,|+\rangle$};
    \node[f4/flag label,anchor=east] at (-0.14,\fFourFlagY) {$\mathscr{S}_{\mathrm{flag}}:\,|0\rangle$};
    \foreach \column/\target in {1/3,2/8,3/4,4/5,5/8,6/6} {
      \pgfmathsetmacro{\fFourGateX}{\fFourGateFirst+(\column-1)*\fFourGatePitch}
      \pgfmathsetmacro{\fFourTargetY}{\fFourCircuitTop-\target*\fFourRowPitch}
      \draw[f4/gate] (\fFourGateX,\fFourSyndromeY) -- (\fFourGateX,\fFourTargetY);
      \fill[f4ink] (\fFourGateX,\fFourSyndromeY) circle[radius=\fFourControlRadius];
      \pic at (\fFourGateX,\fFourTargetY) {f4 target};
    }
    \pic at (\fFourHColumn,\fFourSyndromeY) {f4 hadamard};
    \pic at (\fFourMeasureColumn,\fFourSyndromeY) {f4 meter};
    \pic at (\fFourMeasureColumn,\fFourFlagY) {f4 meter};

    \ifnum\FigFourShowLegend=1\relax
      \begin{scope}[xshift=\fFourLegendLeft cm,yshift=\fFourLegendTop cm]
        \draw[f4rule,line width=0.45pt,rounded corners=0.10cm]
          (0,-3.46) rectangle (7.81,0);
        \draw[f4rule,line width=0.35pt] (0.22,-2.08) -- (7.59,-2.08);

        \filldraw[fill=f4blue,draw=f4blueedge,line width=0.5pt]
          (0.36,-0.32) circle[radius=0.105];
        \node[f4/key] at (0.56,-0.40) {\textbf{W(V)}, $S=1/2$};
        \node[f4/subkey] at (0.56,-0.75) {$[\mathrm{W(CN)_8}]^{3-}$ core};

        \filldraw[fill=f4gold,draw=f4goldedge,line width=0.5pt]
          (0.36,-1.02) circle[radius=0.105];
        \node[f4/key] at (0.56,-1.10) {\textbf{Ln(III)}, capped arm};
        \node[f4/subkey] at (0.56,-1.45) {local / flag ancillae};

        \filldraw[fill=f4outer,draw=f4outeredge,line width=0.5pt]
          (0.36,-1.78) circle[radius=0.105];
        \node[f4/key] at (0.56,-1.86) {\textbf{M}, outer high-spin site};

        \draw[f4/core bond] (4.36,-0.32) -- (4.81,-0.32);
        \node[f4/key] at (4.99,-0.40) {$J'$};
        \node[f4/subkey] at (4.99,-0.75)
          {$\mathrm{W}{-}\mathrm{C}{\equiv}\mathrm{N}{-}\mathrm{Ln}$};

        \draw[f4/outer bond] (4.36,-1.02) -- (4.81,-1.02);
        \node[f4/key] at (4.99,-1.10) {$J$};
        \node[f4/subkey] at (4.99,-1.45) {outer bridge};

        \node[anchor=base west,text=f4subtle,
          font=\normalfont\bfseries\fontsize{8.4}{10.5}\selectfont]
          at (0.30,-2.43) {Alternative choices for M};
        \node[f4/variant] at (0.30,-2.83)
          {$\mathrm{M}=\mathrm{Cr^{III}}$: $\mathrm{Ln}{-}\mathrm{N}{\equiv}\mathrm{C}{-}\mathrm{Cr}(L')$, $S=3/2$};
        \node[f4/variant] at (0.30,-3.24)
          {$\mathrm{M}=\mathrm{Gd^{III}}$: $\mathrm{Ln}{-}\mathrm{F}{-}\mathrm{Gd}(L')$, $S=7/2$};
      \end{scope}
    \fi
  \end{scope}
\fi

\end{tikzpicture}%
\endgroup%